\documentclass[epj]{svjour}
\usepackage{epsfig}
\usepackage{times}
\usepackage{rotating}

\newcommand{\half}{\mbox{${\textstyle \frac{1}{2}}$}}
\newcommand{\reaction}{\mbox{$pp\to dK^+\bar{K}^0$}}
\def\fmn#1#2{\mbox{${\textstyle \frac{#1}{#2}}$}}
\newcommand{\dd}{\mbox{\rm d}}

\begin{document}

\title{Interpretation of $K^{+}\bar{K}^{0}$ pair production in $pp$ collisions}

\author{
  A.~Dzyuba\inst{1,2}%\thanks{email: a.dzyuba@fz-juelich.de}
  \and
  M.B\"uscher\inst{1}\and
  C.~Hanhart\inst{1}\and
  V.~Kleber\inst{3} \and
  V.~Koptev\inst{2}\and
  H.~Str\"{o}her\inst{1}\and
  C.~Wilkin\inst{4}\thanks{email: cw@hep.ucl.ac.uk}
}

\authorrunning{A.~Dzyuba {\it et al.}}

\institute{Institut f\"ur Kernphysik, Forschungszentrum
J\"ulich, 52425 J\"ulich, Germany \and
Petersburg Nuclear Physics Institute, 188300 Gatchina, Russia \and
Physikalisches Institut, Universit\"at Bonn, 53115
Bonn, Germany \and
%
%Institute for Nuclear Research, 117312 Moscow, Russia \and
%
Physics and Astronomy Department, UCL, London WC1E 6BT, UK}

\date{Received: \today}

\abstract{A combined analysis of the published data
on the $pp\to dK^+\bar{K}^{0}$ reaction at excess energies of
47.4\,MeV and 104.7\,MeV is presented. Evidence is found for both
the production of the $a_0^+(980)$ scalar resonance and for a
strong $\bar{K}^0d$ final state interaction.
\PACS{{13.60.Le}{Meson production}\and
{13.75.Jz}{Kaon-baryon interactions}\and
{14.40.Cs}{Other mesons with $S=C=0$, mass $< 2.5\,$GeV}}
}

\maketitle

%
%%%%%%%%%%%%%%%%%%%%%%%%%%%%%%%%%%%%%%%%%%%%%%%%%%%%%%%%%%%%%%%%%%%%%%%%
%
\section{Introduction}
\label{Introduction} The nature of the light scalar resonances
$a_0(980)$ and $f_0(980)$ is still far from being 
understood~\cite{Klempt}. One reason
for this is the lack of precise information about their coupling
to the hadronic channels and especially to the $K\bar{K}$ final
states~\cite{Baru05}. It has been argued that the knowledge of
the couplings might allow one to establish whether these light
scalars are genuine $q\bar{q}$-mesons, or four-quark, or molecular
states~\cite{evidence,Bugg06}. In the case of the $a_0(980)$
resonance, many precise measurements have been performed to
explore the $\pi\eta$-channel. On the other hand, the data
available for the $K\bar{K}$ channel are quite limited, with
large statistical errors and poor mass resolution. Further
experimental data are therefore necessary in order to clarify the
current situation and this was the motivation for the study of
$a_0^+(980)$ production in the $pp\to da_0^+\to dK^+\bar{K}^0$
reaction.

However, kaon pairs can also be created through the $pp\to
K^{+}pY^{0*}$ reaction, where the hyperon decays
\emph{via} $Y^{*}\to \bar{K}^{0}n$. A final state interaction
between the produced neutron and proton can then lead to the
deuteron observed in the $pp\to dK^+\bar{K}^0$ reaction. There
are, of course, several excited hyperons that could contribute to
such a process. Of particular interest for low energy production
are the $\Sigma(1385)$ and the $\Lambda(1405)$. Although the
central values of their masses are below the $\bar{K}N$ threshold,
their widths extend well above, and these hyperons will certainly influence
spectra through a $\bar{K}N$ final state interaction.

There is an extensive program at the COSY COoler SYnchrotron of
the Forschungszentrum J\"ulich to study the production of the
$a_0(980)$ and $f_0(980)$ resonances, as well as the
$\Sigma(1385)$ and $\Lambda(1405)$ hyperons, in nucleon-nucleon
collisions. Using the ANKE magnetic spectrometer and its
associated detector systems, placed at an internal target position
of the storage ring, positive kaons can be identified against a
very high pion and proton background with the help of dedicated
$\Delta E$-$E$ telescopes~\cite{k_nim}. These allow coincidence measurements with
other charged particles to be performed.

The $pp\to dK^+\bar{K}^0$ reaction was investigated at the two
proton beam energies of $T_p{=}2.65$\,GeV~\cite{Kleber04} and
2.83\,GeV~\cite{Dzyuba06}, corresponding to the excess
energies of $\epsilon=47.4$\,MeV and 104.7\,MeV, respectively. In both cases
the $K^+$ and the deuteron were measured directly, with the
$\bar{K}^0$ being identified through a missing-mass peak.

At energies close to the reaction threshold, only a limited number
of partial waves can contribute in the final state. However,
conservation laws demand that there be at least one $p$-wave in
the $dK^+\bar{K}^0$ system. This requirement can be expressed in
one of several different but equivalent coupling schemes. The data
analysis presented in Refs.~\cite{Kleber04,Dzyuba06} considered an
$s$-wave in the $K\bar{K}$ system in association with a $p$-wave
of the deuteron with respect to the meson pair,
\{$(K^+\bar{K}^0)_{s}d$\}$_{p}$, and a $p$-wave $K^+\bar{K}^0$
pair being in an $s$-wave with respect to the deuteron,
\emph{viz.}\ \{$(K^+\bar{K}^0)_{p}d$\}$_{s}$.

It should be noted that the $a_0^+(980)$ can contribute only to
the \{$(K^+\bar{K}^0)_{s}d\}_{p}$ configuration, and the separate
analyses of the two data sets revealed the dominance of this
channel at both energies. However, when these results are
transformed into a basis of the type
$\{\{\bar{K}^0d\}_{\ell}K^+\}_{\ell'}$, it is seen that the
$\{\bar{K}^0d\}$ system is also dominantly in the $s$-wave, with
the $K^+$ being in the $p$-wave. This suggests that the data might
be influenced simultaneously by the production of the $a_0^+(980)$
and the $\bar{K}^0d$ final state interaction.

The relationships between the amplitudes and observables in the
different coupling schemes are discussed in Sec.~\ref{sec2}. A
combined fit to the data at the two energies is presented in
Sec.~\ref{sec3}, where it is assumed that the relative strengths
of the amplitudes are constant in Q, apart from the angular momentum
factors. It is to be expected that the shapes of the corresponding
spectra would be distorted by both the $a_0^+(980)$ resonance, and
the $\bar{K}^0d$ final state interaction and the theoretical forms
of these are the subject of Sec.~\ref{sec4}. The possible
influence of these two distortions on the $pp\to dK^+\bar{K}^0$
data is studied in Sec.~\ref{sec5}, where it is suggested that 
both contribute significantly to the total cross
section, though their effects tend largely to mask each other in
the $K^+\bar{K}^0$ mass spectrum. On the other hand, the ratio of
the $\bar{K}^0d$ to $K^+d$ mass distributions seems to depend
primarily on the $\bar{K}^0d$ final state interaction and only
weakly on the parameters of the $a_0^+(980)$ resonance. Our
conclusions and suggestions for further work are outlined in
Sec.~\ref{sec6}.
%
%%%%%%%%%%%%%%%%%%%%%%%%%%%%%%%%%%%%%%%%%%%%%%%%%%%%%%%%%%%%%%%%%%%%%%%%
%
\section{Amplitudes and observables}
\label{sec2}

Since the available data~\cite{Kleber04,Dzyuba06} were taken quite
close to threshold, we here analyse them in terms of the
lowest allowed partial waves. The application of angular momentum
and parity conservation laws, together with the Pauli principle in
the initial state, shows that the final particles in the
\reaction\ reaction cannot all be in relative $s$-states. The
first permitted final states are therefore $Sp$ and $Ps$, where
the first label denotes the orbital angular momentum between the
$K^+\bar{K}^0$ pair and the second that of the pair relative to
the deuteron. In either case the initial $pp$ pair must have
spin-one and be in an odd partial wave. The most general form of
the reaction amplitudes corresponding to these transitions are
then:
\begin{eqnarray}
\nonumber \mathcal{M}_{Sp} &=&
a_{Sp}(\hat{\vec{p}}\cdot\vec{S})(\vec{k}\cdot\vec{\epsilon^{\dagger}})
+b_{Sp}(\hat{\vec{p}}\cdot\vec{k})(\vec{S}\cdot\vec{\epsilon^{\dagger}}) \\
&&\hspace{-1.2cm}+c_{Sp}(\vec{k}\cdot\vec{S})(\hat{\vec{p}}\cdot\vec{\epsilon^{\dagger}})+
d_{Sp}(\hat{\vec{p}}\cdot\vec{S})(\hat{\vec{p}}\cdot\vec{\epsilon^{\dagger}})
(\vec{k}\cdot\hat{\vec{p}}). \label{eq:M_Sp}\\ %
\nonumber \mathcal{M}_{Ps} &=&
a_{Ps}(\hat{\vec{p}}\cdot\vec{S})(\vec{q}\cdot\vec{\epsilon^{\dagger}})
+b_{Ps}(\hat{\vec{p}}\cdot\vec{q})(\vec{S}\cdot\vec{\epsilon^{\dagger}})\\
&&\hspace{-1.2cm}+c_{Ps}(\vec{q}\cdot\vec{S})(\hat{\vec{p}}\cdot\vec{\epsilon^{\dagger}})+
d_{Ps}(\hat{\vec{p}}\cdot\vec{S})(\hat{\vec{p}}\cdot\vec{\epsilon^{\dagger}})
(\vec{q}\cdot\hat{\vec{p}}). \label{eq:M_Ps}
\end{eqnarray}

Here $\vec{q}$ is the relative momentum between the two kaons,
$\vec{k}$ the deuteron cms momentum, and $\hat{\vec{p}}$ is the
unit vector parallel to the beam direction, also in the
centre-of-mass system. The polarisation vectors of the initial
$pp$ pair and the final deuteron are denoted by $\vec{S}$ and
$\vec{\epsilon}$, respectively. The two sets of coefficients $a$,
$b$, $c$ and $d$ are independent complex amplitudes which may, in
principle, depend upon the scalar kinematic quantities $q^2$,
$k^2$, and $\vec{k}\cdot\vec{q}$.

Since no spin dependence has yet been measured in the \reaction\
reaction, the square of the matrix element must be averaged over
the initial and summed over the final spins. This leads to an
expression in terms of the three-momenta that characterise the
system:
\begin{eqnarray}
\overline{\vert\mathcal{M}(\vec{k},\vec{q})\vert ^2} &=&
C_{0}^{k}k^{2} + C_{1}(\hat{\vec{p}}\cdot\vec{k})^{2} +
C_{0}^{q}q^{2} + C_{2}(\hat{\vec{p}}\cdot\vec{q})^{2} \nonumber\\
&&\hspace{5mm}+\:C_{3}(\vec{k}\cdot\vec{q}) +
C_{4}(\hat{\vec{p}}\cdot\vec{k})(\hat{\vec{p}}\cdot\vec{q})\,.
\label{ansatz}
\end{eqnarray}

The $C$-coefficients are bilinear combinations of the amplitudes
of Eqs.~(\ref{eq:M_Sp}) and (\ref{eq:M_Ps})~\cite{report}:
\begin{eqnarray}
\nonumber%\label{eq:C0q}
C_{0}^{q} &=& \fmn{1}{2}(\vert a_{Ps}\vert^{2}+\vert c_{Ps}\vert^{2}),\\
\nonumber%\label{eq:C0k}
C_{0}^{k} &=& \fmn{1}{2}(\vert a_{Sp}\vert^{2}+\vert c_{Sp}\vert^{2}),\\
\nonumber%\label{eq:C1}
C_{1} &=& \vert b_{Sp}\vert^{2}+\fmn{1}{2}\vert
b_{Sp}+d_{Sp}\vert^{2}\\ \nonumber
&&+Re\left\{a^{*}_{Sp}c_{Sp}+(a_{Sp}+c_{Sp})^{*}(b_{Sp}+d_{Sp})\right\},\\
\nonumber%\label{eq:C2}
C_{2} &=& \vert b_{Ps}\vert^{2}+\fmn{1}{2}\vert
b_{Ps}+d_{Ps}\vert^{2}\\ \nonumber
&&+Re\left\{a^{*}_{Ps}c_{Ps}+(a_{Ps}+c_{Ps})^{*}(b_{Ps}+d_{Ps})\right\},\\
\nonumber%\label{eq:C3}
C_{3} &=& Re\left\{a_{Ps}a^{*}_{Sp}+c_{Ps}c^{*}_{Sp}\right\},\\
\nonumber C_{4} &=& Re\left\{(a_{Ps}+c_{Ps}+b_{Ps}+d_{Ps})^{*}\right.\\
\label{eq:C4}
&&\left.\times(a_{Sp}+c_{Sp}+b_{Sp}+d_{Sp})+2b_{Ps}^{*}b_{Sp})\right\}.
\end{eqnarray}

It is seen from Eqs.~(\ref{eq:C4}) that only $K\bar{K}$ $p$-waves
contribute to $C_{0}^{q}$ and $C_{2}$ and only $s$-waves to
$C_{0}^{k}$ and $C_{1}$. The coefficients $C_{3}$ and $C_{4}$
represent the $s$--$p$ interference terms. In a first
approximation it might be assumed that close to threshold the $C$
are constant, and the data were analysed on the basis of this 
\emph{ansatz}~\cite{Kleber04,Dzyuba06}. However, whereas the
spectra obtained at the two different energies could be fitted
well individually, a combined analysis of both data sets was not
possible while keeping the assumption that the aforementioned
coefficients are constant. 

The coefficients $C$
must be such that the absolute square of the matrix element of
Eq.~(\ref{ansatz}) should not be negative anywhere in the allowed
$(\vec{k},\vec{q})$ space. In practice it is sufficient to impose
the somewhat weaker condition:
\begin{equation}
\label{condition}%
4(C_0^k+\beta\,C_1)(C_0^q+\beta\,C_2)\geq \left(C_3 + \beta\,
C_4\right)^2
\end{equation}
for all $\beta$ in the range $0 \leq \beta\leq 1$. This positivity
requirement was not met by the earlier analyses~\cite{Kleber04,Dzyuba06}.

The presence of the $a_0^+(980)$ could distort the $K^+\bar{K}^0$
$s$-wave amplitudes of Eq.~(\ref{eq:M_Sp}) and hence introduce a
momentum dependence into some of the $C$ coefficients of
Eq.~(\ref{ansatz}). However, this basis is not optimal if the
variation with momentum is due mainly to a final-state interaction
between the $\bar{K}^0$ and the deuteron. For this it is better to
couple first the $\bar{K}^0d$ system, \emph{i.e.}\ use the basis
$\left\{\{\bar{K}^0d\}_lK^+\right\}_{l'}$. The relationship between
the amplitudes in the two bases is easily found by defining
$\vec{Q}$ as the relative $\bar{K}^0d$ momentum and $\vec{K}$ as
the momentum of the $K^+$ in the overall cms.
Non-relativistically, these are connected to the original momenta
through
\begin{eqnarray}
\vec{k} &=& \vec{Q} - \alpha\vec{K},\nonumber\\
\vec{q} &=& \left(1-\fmn{1}{2}\alpha\right)\vec{K}+\vec{Q},
\label{eq:transform}
\end{eqnarray}
where the kinematic factor $\alpha = m_{d}/(m_{d}+m_K)$ depends
upon the deuteron and kaon masses~\cite{report}. The desired expressions are
then found by inserting Eq.~(\ref{eq:transform}) into
Eqs.~(\ref{eq:M_Sp}) and (\ref{eq:M_Ps}).

The spin-averaged matrix element squared has exactly the same
general structure as that of Eq.~(\ref{ansatz}) but in the new
variables:
\begin{eqnarray}
\overline{\vert\mathcal{M}(\vec{K},\vec{Q})\vert ^2} &=&
B_{0}^{K}K^{2} + B_{1}(\hat{\vec{p}}\cdot\vec{K})^{2} +
B_{0}^{Q}Q^{2} + B_{2}(\hat{\vec{p}}\cdot\vec{Q})^{2} \nonumber\\
&&\hspace{5mm}+\:B_{3}(\vec{K}\cdot\vec{Q}) +
B_{4}(\hat{\vec{p}}\cdot\vec{K})(\hat{\vec{p}}\cdot\vec{Q})\,.
\label{ansatz2}
\end{eqnarray}

The interpretation is also completely analogous, with $B_{0}^{K}$
and $B_{1}$ representing the $\bar{K}^0d$ $s$-waves and
$B_{0}^{Q}$ and $B_{2}$ the $p$-waves. The transformation of the
observables between the two bases is given by
\begin{eqnarray}
B_{0}^{Q} &=& C_{0}^{k} + \fmn{1}{4}C_{0}^{q} + \fmn{1}{2}C_{3},\nonumber\\
B_{2} &=& C_{1} + \fmn{1}{4}C_{2} + \fmn{1}{2}C_{4} ,\nonumber\\
B_{0}^{K} &=& \alpha^{2}C_{0}^{k}+\fmn{1}{4}(2-\alpha)^{2}C_{0}^{q} - \fmn{1}{2}\alpha(2-\alpha)C_{3}, \nonumber\\
B_{1} &=& \alpha^{2}C_{1} + \fmn{1}{4}(2-\alpha)^{2}C_{2} - \fmn{1}{2}\alpha(2-\alpha)C_{4},\nonumber\\
B_{3} &=& -2\alpha C_{0}^{k}+\fmn{1}{2}(2-\alpha)C_{0}^{q} + (1-\alpha)C_{3},\nonumber\\
B_{4} &=& -2\alpha
C_{1}+\fmn{1}{2}(2-\alpha)C_{2}+(1-\alpha)C_{4}\,.
\label{transformation}
\end{eqnarray}

\begin{figure*}[htb]
 \centering
 \includegraphics[width=12.cm]{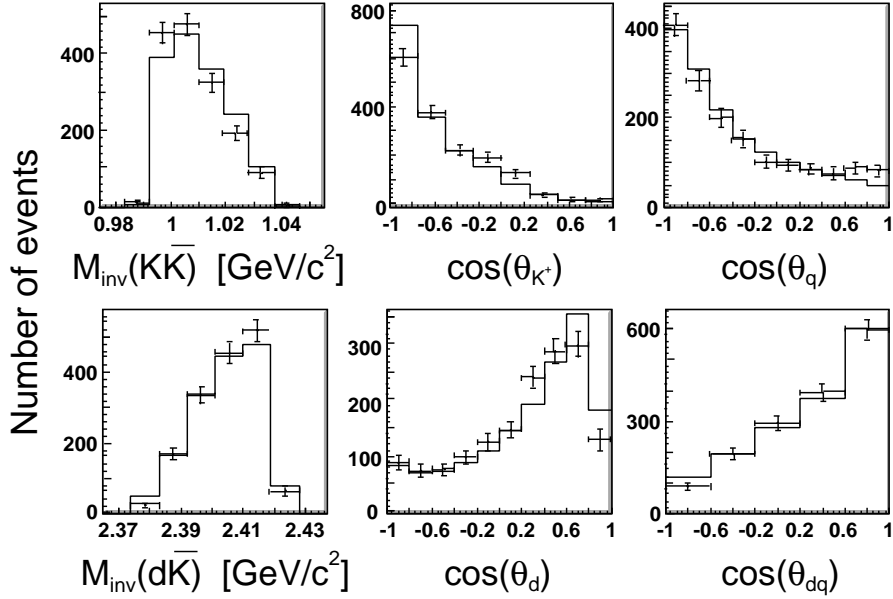}
\caption{Best combined fit to the efficiency-corrected numbers of
events in the 47.4\,MeV  data  
on the basis of the constant
amplitude \emph{ansatz} of Eq.~(\ref{ansatz}). Here
$(\theta_{K^+},\theta_d,\theta_q)$ are the c.m.\ angles of the
$K^+$, deuteron, and $K^+\bar{K}^0$ relative momentum with respect
to the proton beam direction. The angle between the $K^+\bar{K}^0$
relative momentum and the final deuteron momentum is denoted by
$\theta_{dq}$.  \label{fig1}       }
\end{figure*}

\begin{figure*}[htb]
 \centering
 \includegraphics[width=12.cm]{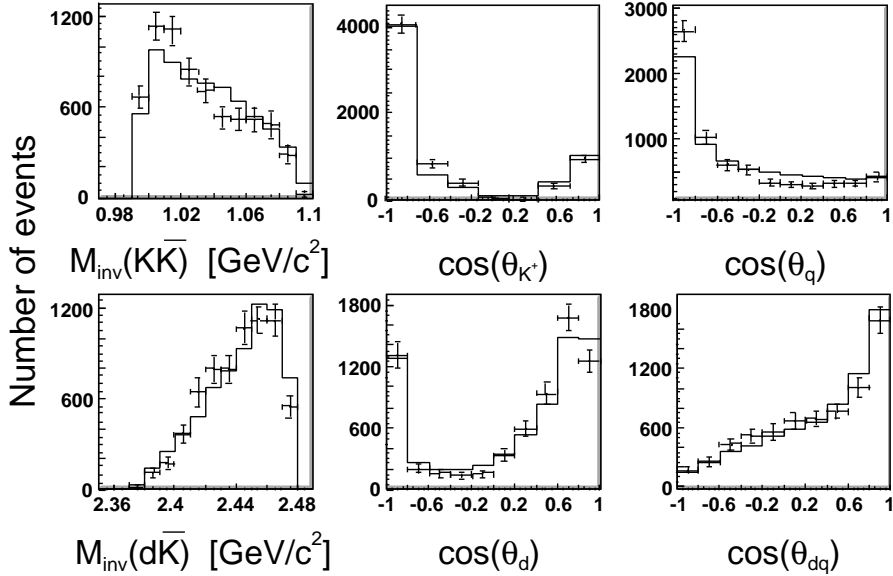}
\caption{The same as Fig.~1 for the 104.7~MeV data.}\label{fig2}
\end{figure*}

It should be noted that if the positivity conditions of
Eq.~(\ref{condition}) are imposed in the
$\left\{\{\bar{K}^0K^+\}d\right\}$ basis, they are automatically
satisfied in the $\left\{\{\bar{K}^0d\}K^+\right\}$ basis, and
\textit{vice-versa}.

The amplitudes are normalised such that, for a fixed total c.m.\
energy $\sqrt{s}$, the total cross section is given by
\begin{equation}
\sigma = \frac{1}{64\pi^3s\,p_p}\int^{\sqrt{s}-m_{3}}_{m_{1}+m_{2}}
\,p_3\,p_1^*\,
\langle\,\overline{\vert\mathcal{M}^{2}}\vert\,\rangle\,\dd M_{12}\,,
\label{sigtot}
\end{equation}
where $\langle\,\overline{\vert\mathcal{M}^{2}}\vert\,\rangle$ is
the angular-average of the squared transition matrix element of
Eq.~(\ref{ansatz}) or (\ref{ansatz2}). The momentum $p_p$ of the
incident proton and $p_3$ of particle-3 are evaluated in the
overall c.m.\ frame, while that of particle-1, $p_1^*$, is
considered in the rest frame of the $(12)$ system, where the
effective mass is $M_{12}$.
%
%%%%%%%%%%%%%%%%%%%%%%%%%%%%%%%%%%%%%%%%%%%%%%%%%%%%%%%%%%%%%%%%%%%%%%%%
%
\section{Solutions with constant coefficients}
\label{sec3} \setcounter{equation}{0}

In order to determine the $C$ parameters defined in
Eq.~(\ref{ansatz}), we performed fits of the measured observables
using GEANT simulated data samples. This method, which is
described in detail in Ref.~\cite{Dzyuba06}, works directly with
the distributions uncorrected for acceptance, which is then taken
into account in the simulation. The results of making separate
single-energy fits, as well as of a combined study of the two
energies, are shown in Table~1.

\begin{table}[hbt]
\begin{center}
\caption{Results of the fits to the shapes of the differential 
distributions on the basis of Eq.~(\ref{ansatz}) with constant 
coefficients. The parameters are measured with respect to
$C^{k}_{0}$, whose value is put equal to unity. The fits push
$C^{q}_{0}$ and $C_{3}$ to the limits allowed by
Eq.(\ref{condition}) and so they have here both been fixed at
zero. The probabilities $P$ of the different partial-wave
configurations have typically uncertainties of $\pm5\%$.}
\label{tab:single}
\begin{tabular}{c | c | c | c}
$Q$ [MeV] & 47.4 & 104.7 & combined\\
\hline
\hline
$C^{q}_{0}$ & 0    & 0    & 0 \\
$C^{k}_{0}$ & 1    & 1    & 1 \\
$C_{1}$     & $0.2^{+0.33}_{-0.28}$  & $2.5^{+1.8}_{-0.95}$  & $0.83^{+0.34}_{-0.24}$\\
$C_{2}$     & $1.1^{+0.39}_{-0.34}$  & $3.9^{+2.2}_{-1.15}$  & $1.95^{+0.28}_{-0.30}$\\
$C_{3}$     & 0    & 0    & 0\\
$C_{4}$     & $-1.9^{+0.44}_{-0.50}$ & $-6.4^{+1.6}_{-3.1}$ & $-3.30^{+0.38}_{-0.44}$ \\
\hline
$P\left(\{\{K^+\bar{K}^0\}_{s}d\}_{p}\right)$ & 80\% & 79\% & 84\%\\
$P\left(\{\{\bar{K}^0d\}_{s}K^+\}_{p}\right)$ & 64\% & 72\% & 65\%\\
$P\left(\{\{K^+d\}_{s}\bar{K}^0\}_{p}\right)$ & 23\% & 16\% & 21\%\\
\hline
$\chi^{2}/$ndf & 74.5/42 & 126/55 & 251/100\\
\end{tabular}
\end{center}
\end{table}

The current fits include the positivity constraints expressed by
Eq.~(\ref{condition}), which were not taken into account in the
previous analyses~\cite{Kleber04,Dzyuba06}. Fortunately, the
influence of these constraints on the observables is not very
large and the main results are largely unaffected.

The lowest value of $\chi^2$ would be achieved at both energies
with a negative $C^{q}_{0}$ coefficient. As shown in
Eq.~(\ref{eq:C4}), $C^{q}_{0}$ is given by absolute squares of amplitudes
that cannot be negative. We therefore set $C^{q}_{0}=0$ in the fit
which, as a consequence of Eq.~(\ref{condition}), means that
$C_{3}$ must also vanish. The results of the combined fit to the
numbers of events for different observables, corrected for chamber
efficiencies, are shown in Figs.1 and 2 for the two energies.

The most obvious and striking of the results reported in
Table~\ref{tab:single} are the probabilities for the different
partial-wave configurations. The $K^{+}\bar{K}^{0}$ subsystem is
overwhelmingly in the $s$-wave, with only about 15\% being in a
$p$-wave. It is therefore clear that the $a_0^{+}(980)$ channel is
strongly favoured in this reaction. However, it is equally evident
that, since the $s$-wave probability for $\bar{K}^0d$ of about two
thirds is three times that for $K^+d$, the antikaon is also
strongly attracted to the baryons.

\begin{figure}[htb]
\vspace*{3mm}
\centering
\includegraphics[width=8.cm]{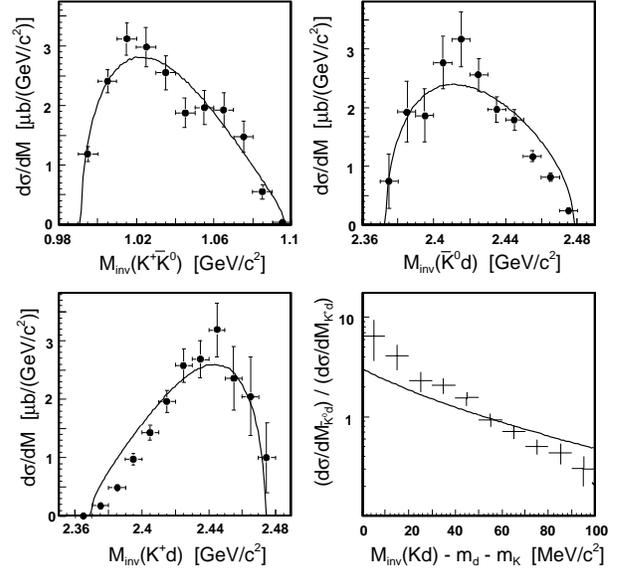}
\caption{Invariant mass distributions of the $pp\to
dK^+\bar{K}^{0}$ differential cross sections at 104.7\,MeV. These
are compared with the constant coefficient solution of the
combined fit whose parameters are given in Table~1. Also shown is
the ratio of the $\bar{K}^0d$ and $K^+d$ distributions, plotted as
a function of the  mass excess $\Delta M =
M_{\textrm{inv}}(Kd)-m_K-m_d$.} \label{fig:mass}
\end{figure}

As a cross-check on the analysis, the shapes for the distributions
obtained with the best global fit parameters were compared with
those published in Ref.~\cite{Dzyuba06} and reasonable agreement
was found. Since the previous analysis was done on the basis of
acceptance-corrected data, the accord means that there is little
ambiguity in the results originating from the uncertainties in the
very non-uniform ANKE acceptance.

Figure~\ref{fig:mass} shows the differential cross sections in
terms of the three two-particle invariant masses for the
104.7\,MeV data compared to the results of the global fit. These
acceptance-corrected $\bar{K}^0d$ and $K^+d$ distributions are
very different, with lower masses being strongly favoured in the
former case. This is even more obvious if we compare the ratio of
the $\bar{K}^0d$ to $K^+d$ distributions which, to eliminate the
effects of the kaon mass difference, is shown in
Fig.~\ref{fig:mass} as a function of the mass excess $\Delta M =
M_{\textrm{inv}}(Kp)-m_K-m_p$. The ratio falls very fast with
$\Delta M$ in a way that is very similar to that found for the
ratio of the $K^-pp$ and $K^+pp$ invariant mass distributions in
the $pp\to ppK^+K^-$ reaction~\cite{Yoshi}. In this case the
strong variation could be explained in terms of a $K^-p$ final
state interaction, perhaps driven by the $\Lambda(1405)$, though
it should be noted that the angular momentum constraints are far
weaker there.

Although the overall description of the $pp\to dK^+\bar{K}^{0}$
results shown in Fig.~\ref{fig:mass} is reasonable, it is clear
that the data are lower than the fits for high $\bar{K}^0d$ and
low $K^+d$ masses and this is particulary striking in the plot of
their ratio. A more satisfactory description of these data would
be achieved with a somewhat larger $s$-state $\bar{K}^0d$
probability, as indicated by the single energy fit shown in
Table~\ref{tab:single}. However, the disagreement may also be a
reflection of the distortions introduced by the $a_0^+(980)$
resonance and the $\bar{K}^0d$ final state interaction, to which
we now turn.
%
%%%%%%%%%%%%%%%%%%%%%%%%%%%%%%%%%%%%%%%%%%%%%%%%%%%%%%%%%%%%%%%%%%%%%%%%
%

\section{Influence of the $\mathbf{a_0^+(980)}$ resonance and
$\mathbf{\bar{K}^{0}d}$ interaction in final state}%
\label{sec4} \setcounter{equation}{0}%

Because the mass of the $a_0(980)$ resonance is very close to the
$K\bar{K}$ threshold, its shape is not described satisfactorily by
a Breit-Wigner form. The energy dependence of the width is taken
into account in the standard  Flatt\'e parameterisation for the
propagator~\cite{Flatte76}:

\begin{equation}
G_{a_{0}}(q_{K\bar{K}})=\frac{N}{M_{a_{0}}^{2}-m_{KK}^{2}
-iM_{a_{0}}(\bar{g}_{\pi\eta}q_{\pi\eta}+\bar{g}_{K\bar{K}}q_{K\bar{K}})}\,,
\label{eq:a0}
\end{equation}
where $\bar{g}_{\pi\eta}$ and $\bar{g}_{K\bar{K}}$ are
dimensionless coupling constants and very often it is their ratio
$R = \bar{g}_{K\bar{K}}/\bar{g}_{\pi\eta}$ that is
quoted~\cite{Baru05}. For convenience, the normalisation constant
$N$ is chosen here such that $G_{a_{0}}$ is
equal to be unity at the $K^+\bar{K}^0$
threshold. The nominal mass of the resonance is $M_{a_{0}}$,
$m_{K\bar{K}}$ is the $K\bar{K}$ invariant mass, and $q_{\pi\eta}$
is the $\pi\eta$ relative momentum in the $a_0$ rest frame. It is
important to note that the relative momentum in the $K\bar{K}$
channel,
\begin{equation}
q_{K\bar{K}} = \half\sqrt{m_{K\bar{K}}^{2}-4m_{K}^{2}}
\end{equation}
is positive imaginary below the $K\bar{K}$ threshold.

\begin{figure}[htb]
  \centering
\includegraphics[width=7cm]{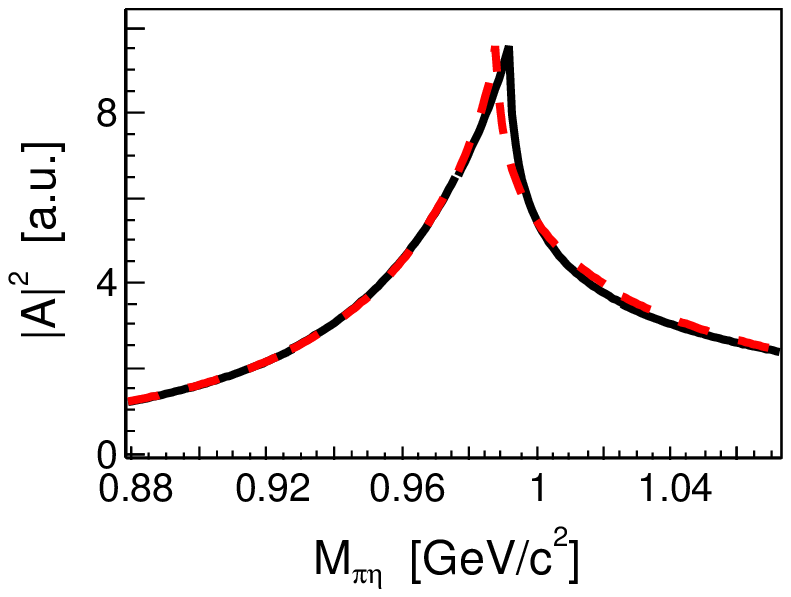}
\includegraphics[width=7cm]{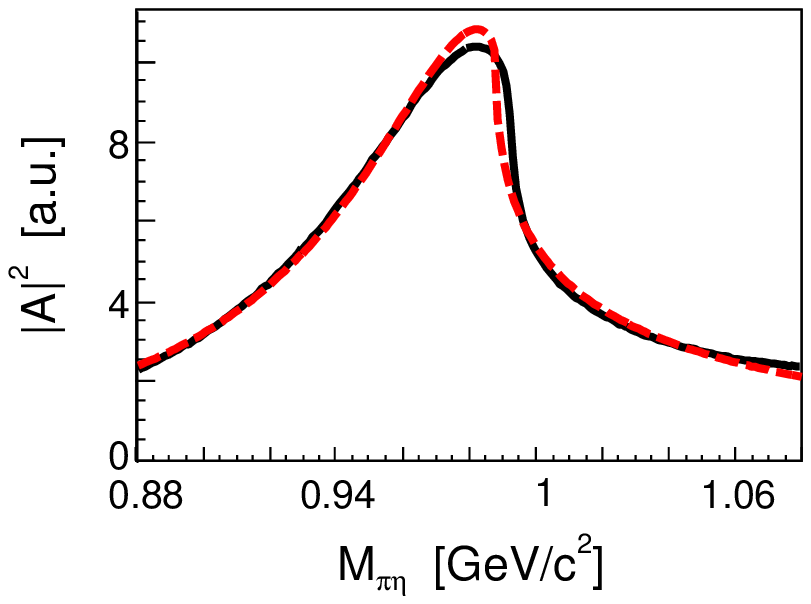}
\caption{Upper panel: The Achasov parameterisation (solid
line)~\cite{Achasov03} compared with a non-relativistic
$K^+\bar{K}^0$ Flatt\'{e} function (dashed line) with the
following parameters: $\bar{g}_{\pi\eta} = 0.614$, $R = 1.650$,
$M_{a_{0}} = 1.020\,$GeV/c$^{2}$. Bottom panel: The same for the
Oller parameterisation~\cite{Oller99} with following parameters:
$\bar{g}_{\pi\eta} = 0.698$, $R = 1.463$, and $M_{a_{0}} =
0.959\,$GeV/c$^{2}$. }
  \label{fig:comparison}
\end{figure}

The uncertainties in the $a_{0}(980)$ parameters as extracted from
the data collected in the different experiments are rather large,
as can be seen from the compilation given in Ref.~\cite{Baru05}.
Although the statistical errors for each experiment are small,
there are significant discrepancies between the quoted values of
the coupling constants. This might indicate either large
systematic uncertainties or be a consequence of a certain scale
invariance of the Flatt\'e parameterisation~\cite{Baru05}. The
picture is, however, little changed if one uses instead a
parameterisation based on more fundamental theoretical
ideas~\cite{Achasov03,Oller99,others}. That of
Achasov~\cite{Achasov03} takes into account the so-called
finite-width corrections to the self-energy loop. As shown in
Ref.~\cite{Baru05}, these different parameterisations are
equivalent in the non-relativistic limit. To illustrate this, we
show  in Fig.~\ref{fig:comparison} that the spectra of both
Ref.~\cite{Achasov03} and Ref.~\cite{Oller99} can be well
described by a non-relativistic Flatt\'e distribution with
modified parameters. With the currently available data it is not
possible to distinguish between the relativistic and
non-relativistic forms.

It is to be expected that the $pp\to dK^+\bar{K}^{0}$ amplitudes
leading to the $s$-wave $K^+\bar{K}^0$ final state should be
modified through the introduction of Flatt\'e propagator
$G_{a_0}$. This effectively introduces a momentum dependence into
some of the amplitudes of Eq.~(\ref{eq:M_Ps}) and hence into the
corresponding observables through Eq.~(\ref{eq:C4}).

There is also significant uncertainty in the low energy $\bar{K}d$
interaction, which is nicely summarised in Ref.~\cite{Rusetsky}.
The $\bar{K}d$ scattering lengths quoted there are typically
\begin{equation}
a_{\bar{K}d} \approx (-1.0+i1.2)\,\textrm{fm}\,,
\end{equation}
though the spread is quite large, depending upon the theoretical
assumptions and the experimental data.

In the scattering length approximation, the amplitude for the
production of an $s$-wave $\bar{K}d$ pair should be multiplied by
an enhancement factor
\begin{equation}
\label{Kdfsi}
F_{\bar{K}d}(Q) = \frac{1}{\left(1-iQa_{\bar{K}d}\right)}\,,
\end{equation}
which depends on the relative $\bar{K}d$ momentum $Q$.
%
%%%%%%%%%%%%%%%%%%%%%%%%%%%%%%%%%%%%%%%%%%%%%%%%%%%%%%%%%%%%%%%%%%%%
%
\section{Fit procedure and results}
\label{sec5} \setcounter{equation}{0}

\begin{figure}[htb]
\vspace*{3mm}
\centering
\includegraphics[width=8.cm]{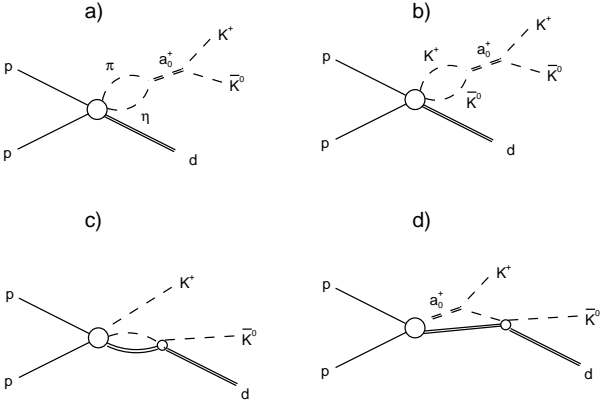}
\caption{Diagrams for $pp{\to}dK^{+}\bar{K}^{0}$ reaction:
a,b) correspond to direct $a_0(980)$ production involving strong
$(\pi\eta,K\bar{K})$ coupled channel effects,
c) reflects non-resonant $dK^{+}\bar{K}^{0}$ production
corrected by the $\bar{K}d$ final state interactions.
d) includes both ``rescattering'' effects.}
\label{fig:feynman}
\end{figure}

With the inclusion of final state interactions in both the
$K^+\bar{K}^0$ and $\bar{K}^0d$ $s$-wave systems, one should study
the contributions indicated symbolically in Fig.~\ref{fig:feynman}
as well as higher order rescatterings.

It is important to stress at the outset that the final state
interaction factors of Eqs.~(\ref{eq:a0}) and (\ref{Kdfsi})
involve unknown overall normalisations and will, at most, only
describe the momentum dependence of the corresponding $s$-wave
observable. In particular, one cannot determine the modifications
to the $s$-wave probabilities induced by the final state
interactions. For this purpose one would need full $K\bar{K}$ and
$\bar{K}d$ ``potentials'' or their equivalent. We therefore limit
ourselves here to the study of how the combined final state
interactions distort the mass distributions of
Fig.~\ref{fig:comparison}. For this we follow the procedure of
Ref.~\cite{Yoshi} and assume that the final state interaction
factors are multiplicative.

In principle, only the $s$-wave amplitudes should be multiplied by
the corresponding final state interaction factor of
Eq.~(\ref{eq:a0}) or Eq.~(\ref{Kdfsi}). However, we have seen from
Table~\ref{tab:single} that the $s$ waves are dominant in both the
$K^+\bar{K}^0$ and $\bar{K}^0d$ channels. We therefore make the
drastic simplification of multiplying the matrix-element-squared
$\langle\,\overline{\vert\mathcal{M}^{2}}\vert\,\rangle$ by the
product of the absolute-squares of the final state interaction
factors:
\begin{equation}
\label{simple}
\langle\,\overline{\vert\mathcal{M}^{2}}\vert\,\rangle \longrightarrow
\langle\,\overline{\vert\mathcal{M}^{2}}\vert\,\rangle \times
|F_{\bar{K}d}(Q)|^2 \times |G_{a_0}(q)|^2\,.
\end{equation}
This \textit{ansatz} means that the $p$-waves are also
modified even though there should be no final state interactions
in these channels. On the other hand, the procedure avoids the
introduction of extra parameters that depend upon the relative
phases of the $s$ and $p$ waves.

\begin{figure}[htb]
\vspace*{3mm}
\centering
\includegraphics[width=8.cm]{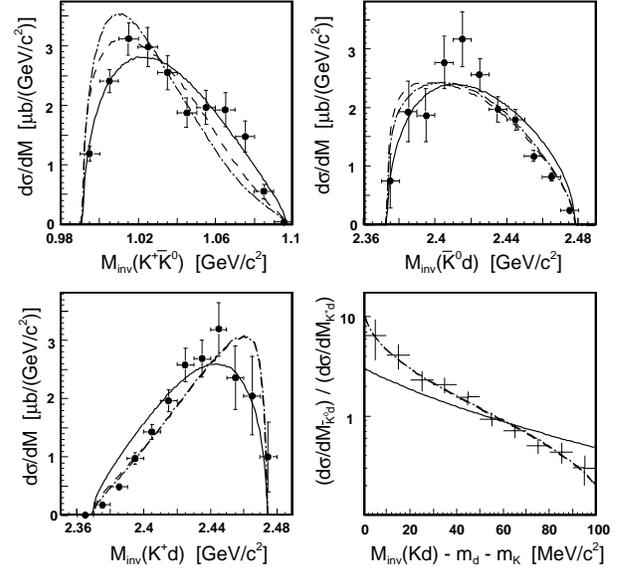}
\caption{Invariant mass distributions of the $pp\to
dK^+\bar{K}^{0}$ differential cross sections at 104.7\,MeV
compared with predictions of the combined fit, where \textit{fsi}
distortions are introduced using the \textit{ansatz} of
Eq.~(\ref{simple}) with the Bugg \textit{et al.}~\cite{Bugg94}
(dashed curve) and Teige \textit{et al.}~\cite{Teige99} (chain
curve) parameterisations of the $a_0^+(980)$. The results of the
constant amplitude approach of Fig.~\ref{fig:mass} are presented
for orientation (solid line). The fit to the ratio of the
$\bar{K}^0d$ and $K^+d$ distributions is insensitive to the
parameters of the $a_0^+(980)$. } \label{fig:mass2}
\end{figure}

A new combined fit to the two data sets was undertaken on the
basis of the \textit{ansatz} of Eq.~(\ref{simple}), and the results
for the invariant mass distributions at 104.7\,MeV are shown in
Fig.~\ref{fig:mass2} for both the Bugg \textit{et
al.}~\cite{Bugg94} and Teige \textit{et al.}~\cite{Teige99}
parameterisations of the $a_0^+(980)$ which correspond,
respectively, to a wide and narrow resonance. The $K^+\bar{K}^0$
spectrum is better described when the wider $a_0^+(980)$ is used
but the major improvement seen from the introduction of these
distortions is in the $\bar{K}^0/K^+$ ratio, where the excellent
description depends relatively little on the parameters chosen for
the $a_0^+(980)$ resonance and almost entirely upon the
$\bar{K}^0d$ \textit{fsi}.

\begin{figure}[htb]
\vspace*{3mm}
\centering
\includegraphics[width=8.cm]{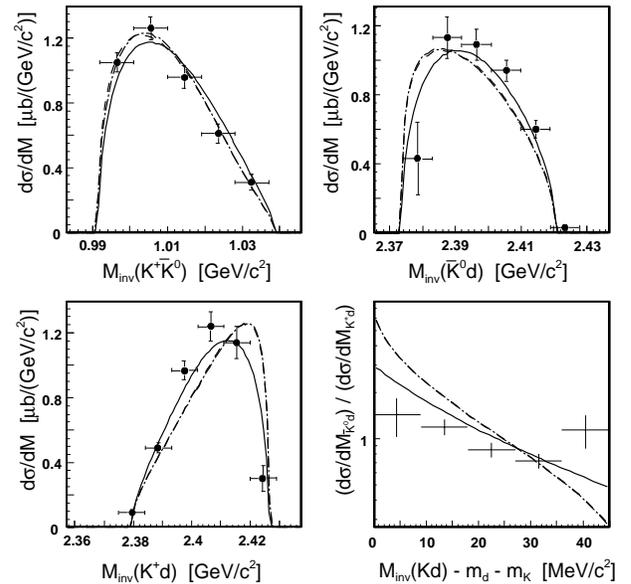}
\caption{As for Fig.~\ref{fig:mass2} but at an excess energy of 47.4\,MeV.}
\label{fig:mass3}
\end{figure}

Owing to the relatively small phase space volume, the data at
47.4\,MeV are not sensitive to the $a_0^+(980)$ parameters
(Fig.~\ref{fig:mass3}). The ratio of the $\bar{K}^0d$ and $K^+d$
distributions is enhanced at the low mass region. However, this
enhancement is weaker than the result of the fit which is based on
Eq.~(\ref{simple}).

\begin{figure}[htb]
\vspace*{3mm}
\centering
\includegraphics[width=6.cm,angle=-90]{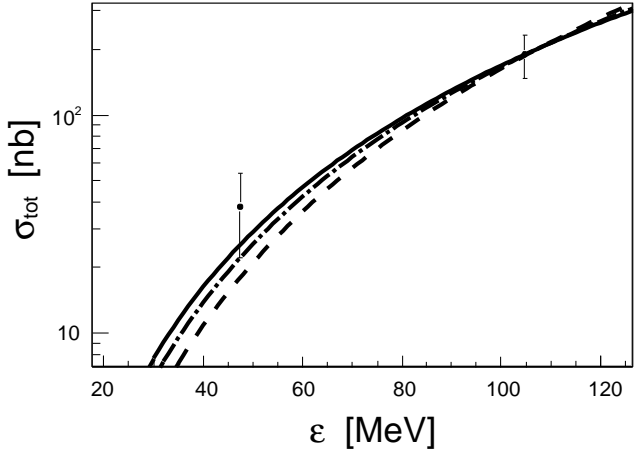}
\caption{Total cross section for the $pp\to dK^{+}\bar{K}^{0}$
reaction as a function of the excess energy $\epsilon$. Points with
errors are the cross sections measured with the ANKE
spectrometer~\cite{Kleber04,Dzyuba06}. The dashed line, which is
normalised on the 104.7\,MeV point, shows a phase space simulation
with the fixed partial wave contributions, as defined by the joint
fit of Table~1. The solid line shows the corresponding energy
dependence after the introduction of the \textit{fsi} in both the
$a_{0}^{+}(980)$ and $\bar{K}d$ channels. The inclusion of either
the $a_{0}^{+}(980)$ or $\bar{K}d$ alone leads to almost identical
results, which are shown by the chain curve.} \label{fig:total_XS}
\end{figure}

At low energies both the $K^+\bar{K}^0$ and $\bar{K}^0d$ systems
must be in regions where the cross section is enhanced by the two
\textit{fsi}. As a consequence, the total cross section predicted
using the \textit{ansatz} of Eq.~(\ref{simple}) shows a slower
energy variation than when the constant coefficients are used.
This is seen clearly in Fig.~\ref{fig:total_XS}, where the results
at the two measured energies~\cite{Kleber04,Dzyuba06} are compared
with predictions that are normalised to the 104.7\,MeV point.

The introduction of either \textit{fsi} increases the cross
section at the lower energy by a similar amount and the inclusion
of both effects improves significantly the agreement with the
47.4\,MeV point.

%
%%%%%%%%%%%%%%%%%%%%%%%%%%%%%%%%%%%%%%%%%%%%%%%%%%%%%%%%%%%%%%%%%%%%%
%
\section{Conclusions}
\label{sec6} \setcounter{equation}{0}

We have presented a combined analysis of the measurements of the
$pp\to dK^+\bar{K}^0$ reaction made with the ANKE spectrometer at
COSY at excess energies of 47.4 and 104.7\,MeV. The application of
the Pauli principle, combined with the conservation of angular
momentum and parity, shows that an overall $s$-wave is forbidden
in the final state. There must be at least one $p$ or higher wave.
The data clearly demonstrate that both the $K^+\bar{K}^0$ and
$\bar{K}^0d$ systems are dominantly in $s$-waves, while $p$-waves
dominate the $K^{+}d$ channel. The big difference between the kaon
and antikaon interactions with the deuteron is seen most clearly
in the ratio of the differential cross sections in terms of the
$Kd$ invariant mass. This ratio seems to depend primarily on the 
$\bar{K}d$ interaction. The fast variation observed there is similar
to that found in the $K^-pp/K^+pp$ ratio measured in the $pp\to
ppK^+K^-$ reaction.

The effects of the $a_0^+(980)$ scalar resonance are more subtle,
though the fact that the $K^+\bar{K}^0$ is almost entirely in an
$s$-wave is a strong indication of its influence. The shapes of
the differential distributions are little changed from those of
the constant parameter solution provided that both the
$\bar{K}^0d$ and $K^+\bar{K}^0$ final state interactions are taken
into account. It is then clear that the $a_0^+(980)$ \textit{fsi}
compensates the $\bar{K}^0d$ distortion of the $K^+\bar{K}^0$
spectrum so that there is a delicate interplay between these two
\textit{fsi}. More theoretical work is clearly necessary here to 
get a fuller understanding and attempts have been made to study
this problem on a more fundamental level~\cite{oom}.

The two final state interactions taken together enhance the ratio
of the total cross section
$\sigma(47.4\,\textrm{MeV})/\sigma(104.7\,\textrm{MeV})$ and lead
to a better description of the published data. 
Within the framework of the simple approach presented here, 
even greater sensitivity would be achieved if we had total cross
section data closer to threshold.

\vspace{0.5cm}

\begin{acknowledgement}
The authors would like to thank many of the other members of the
ANKE collaboration for discussions and especial mention here should go to
V.~Grishina, L.~Kondratyuk, and A.~Sibirtsev.
This work has partially been supported by BMBF, DFG, Russian Academy
of Sciences, and COSY FFE.
\end{acknowledgement}

\end{document}